\documentclass[conference]{IEEEtran}
\IEEEoverridecommandlockouts
\usepackage{cite}
\usepackage{amsmath,amssymb,amsfonts}
\usepackage{algorithmic}
\usepackage{graphicx}
\usepackage{subfig}
\usepackage{url}
\usepackage{textcomp}
\usepackage{xcolor}
\def\BibTeX{{\rm B\kern-.05em{\sc i\kern-.025em b}\kern-.08em
    T\kern-.1667em\lower.7ex\hbox{E}\kern-.125emX}}

\makeatletter
\def\ps@IEEEtitlepagestyle{ 
	\def\@oddfoot{\mycopyrightnotice}
}
\def\mycopyrightnotice{
	{\scriptsize \hspace{-1cm} Presented in International Conference on COMmunication Systems \& NETworkS (COMSNETS) 2022 (MINDS Workshop). Published version: https://doi.org/10.1109/COMSNETS53615.2022.9668495}
}

\begin{document}

\title{Optimal Lockdown to Manage an Epidemic\\
\thanks{This work was funded in part by SERB India via the grant SRG/2019/001054.}
}

\author{\IEEEauthorblockN{Jagtap Kalyani Devendra}
\IEEEauthorblockA{\textit{Department of Electrical Engineering and Computer Science} \\
\textit{Indian Institute of Science Education and Research Bhopal}\\
Madhya Pradesh, 462066, India \\
jagtap19@iiserb.ac.in}
\and
\IEEEauthorblockN{Kundan Kandhway}
\IEEEauthorblockA{\textit{Department of Electrical Engineering and Computer Science} \\
\textit{Indian Institute of Science Education and Research Bhopal}\\
Madhya Pradesh, 462066, India \\
kundankandhway@iiserb.ac.in}
}

\maketitle

\begin{abstract}
We formulate an optimal control problem to determine the lockdown policy to curb an epidemic where other control measures are not available yet. We present a unified framework to model the epidemic and economy that allows us to study the effect of lockdown on both of them together. The objective function considers cost of deaths and infections during the epidemic, as well as economic losses due to reduced interactions due to lockdown. We tune the parameters of our model for Covid-19 epidemic and the economies of Burundi, India, and the United States (the low, medium and high income countries). We study the optimal lockdown policies and effect of system parameters for all of these countries. Our framework and results are useful for policymakers to design optimal lockdown strategies that account for both epidemic related infections and deaths, and economic losses due to lockdown.
\end{abstract}


\section{Introduction}

Epidemics are known to humankind since a long time. Few examples include smallpox, measles, polio, Ebola, Zika etc. Even before Covid-19 era, the world economy suffered an estimated loss of \$570 billion every year due to various epidemic outbreaks \cite{epi_cost}. However, there are few epidemics that have shaped human history as much as Covid-19---estimated losses in United States (US) alone is approximately \$16 trillion \cite{c1_c2_us}. Since the virus was novel, countries had to resort to drastic measure like lockdown to manage the outbreak. The aim of this paper is to formulate an optimal lockdown problem that takes into account both the costs due to infections and deaths, and economic losses due to reduced interaction among the individuals of the population.

Classical literature on epidemic control \cite{b6, b7} employs strategies like vaccination and treatment for suppressing new outbreak of a known epidemic (e.g. rabies). This has limited usefulness for a novel virus like Covid-19. Locking down a city has huge economic implications and costs, as has been witnessed by people throughout the world recently. Authors in \cite{b2, b4} study lockdown policies but without considering the economic implications of locking down the area under study.

A few studies have accounted for economic losses while deciding the lockdown strategies \cite{b1, b3}, however, the economic model is very simple. Specifically, these studies do not consider the impact of reduced number of interaction among individuals of the population on the economy, which is not the case in the present work. The work in \cite{b5} does consider the impact of interactions on economy (and of course the epidemic) but the authors only study heuristic lockdown strategies without formulating any optimization problem.

The contributions of this paper are as follows: (1) We present a unified model for epidemic and economy. Reducing interaction among population (through lockdown) helps prevent epidemic spread, but at the same time reduces economic interactions as well. The model allows us to study this trade-off. (2) We formulate an optimal control problem that takes into account the value of statistical life (deaths due to epidemic), health costs due to infections, economic costs due to loss of economic activities due to lockdown. (3) The proposed model is parameterized based on publicly available Covid-19 and economic data for three countries -- US, India, Burundi -- high, middle and low income economies. Based on this we study optimal lockdown strategies and provide insights to the policymaker.

Based on the calibrated model, the optimal lockdown is partial for US (a high income economy); and almost non-existent for Burundi (a low income economy), for a wide range of `statistical value of human life' for the respective countries. But, for India (a middle income economy) the optimal solution transitions from partial lockdown to full lockdown as the statistical value of human life increases. This is in line with the observation that neither US nor Burundi imposed a nationwide lockdown as a response strategy during first wave of Covid-19 in early 2020. On the other hand, India imposed a nationwide lockdown during the same period \cite{burundi_lock, india_lock, us_lock}.

Rest of this paper is organized as follows. Section II discusses the system model and problem formulation. In Section III we discuss parameter tuning for various countries. Section IV discusses the technique employed to solve the formulated problem. Section V presents the results and finally Section VI concludes the paper.

\section{System Model}

In this section we present an ODE based unified epidemic and economic model. Social interaction contributes to value generation in the economy as well as the spread of disease, the unified framework presented here allows us to study this interdependence. We then present the optimal control problem to obtain the best lockdown policy for countries with different economic conditions.

\subsection{The Epidemic Model} 
We have used the Susceptible-Infected-Recovered-Dead (SIRD) compartmental model for modeling epidemic on homogeneously mixed population. We model a geographical area with bustling economic activity (e.g. a city or capital of the nation). People from other areas migrate to this place at a net migration rate, $\mu$, for seeking employment, tourism, education etc. We consider logistic growth of population with carrying capacity $K$. Logistic growth model for human population has been used extensively in the literature \cite{b5, pop_logistic}. 

The total population of the region of interest is $N(t)$ at any time $t$. Let us assume that epidemic breaks out at time $t = 0$ and ends at $t = T$. The population in the region can be compartmentalized into four compartments/subgroups: susceptible, infected, recovered and dead. Let $S(t)$, $I(t)$, $R(t)$ and $D(t)$ be the respective numbers at time $t$. Therefore, the total population at any time $t$ is $ N(t) = S(t)+I(t)+R(t)$ (dead individuals are not considered in the population). At $t=0$ there are $I(0)$ infected individuals who act as seeds for the spread of the disease.

The spreading rate of the epidemic is given by $\beta$ which is related to the lockdown control signal imposed to curb the disease spread, as discussed in the following. The strength of lockdown at any time $t$ is $l(t) \in [0, l_0], ~l_0<1$. In absence of the epidemic, people interact with an average of $k_{0}$ individuals per unit time. Due to lockdown restrictions, this number reduces to a value $ k = k_{0}(1-l(t))$. As the average number of interactions change, the spreading rate of the epidemic also changes. The spreading rate after imposing the lockdown is given by $\beta = \beta_{0}k$ where $\beta_{0}$ is the disease specific infectivity parameter. Infected individuals recover at a rate of $\gamma$ or the recovery rate (the average number of days in which an infected person recovers is $1/\gamma$). The infected individuals may also die at a rate of $\delta$, the death rate.

The strength of lockdown control signal lies in the range $l(t) \in [0, l_0], ~l_0<1$. In practice, human interactions can never be completely cut-off. Even during the epidemic the essential services (e.g. health care system, grocery etc.) that are necessary for continuation of life keeps running. Hence, $100\%$ lockdown is not feasible. Therefore, the maximum strength of lockdown signal is set to $l_0=0.75$. The dynamics of the epidemic with lockdown as a control signal is given in the following equations.
\begin{subequations}\label{eqn:1}
	\begin{align}
		\dot{S}(t) = \mu S(t) - \frac{\beta S(t)I(t)}{N(t)} - \frac{\mu S(t)N(t)}{K} \\
		\dot{I}(t) = \mu I(t) + \frac{\beta S(t)I(t)}{N(t)} - (\gamma+\delta)I(t) - \frac{\mu I(t)N(t)}{K}  \label{eqn:1b}\\
		\dot{R}(t) = \mu R(t) + \gamma I(t) - \frac{\mu R(t)N(t)}{K}  \label{eqn:1c}\\
		\dot{D}(t) = \delta I(t) \label{eqn:1d}
	\end{align}
\end{subequations}
Notice that during the outbreak, the inflow and outflow of individuals from the region of interest is allowed. These new migrants may fall into any of the three subgroups: susceptible, infected, or recovered. 

\subsection{The Economic Model}
The economy is modeled based on number of economically beneficial interactions among individuals of the population. Each economically beneficial interaction generates a value that accumulates to provide total value to the economy. This is subset of the number of interactions $k=k_{0}(1-l(t))$ every individual makes. Thus, the lockdown that aids in curbing the spread of disease, adversely affects the economy.

Not every social interaction leads to value addition in the economy. As we increase the number of interactions in the population, economically beneficial interactions do not increase linearly, rather it increases sub-linearly. This is because if an individual is forced to make only a few interactions (say, per week), then these will be utilized only for essential reasons. However, as the lockdown eases, the number of purely social interactions, which do not have any economic value, will increase gradually. This effect can be captured by any sub-linear function. For simplicity we have considered a sinusoidal function.

Let $\alpha$ be the fraction of the employed population. Further, let $a_{1}$ be the fraction of interactions that are economically beneficial in absence of any epidemic outbreak. Since any individual interacts with $k_0$ others, the number of economically beneficial interactions in absence of any epidemic is given by 
\begin{equation*}
	\hat{n}_{ei}(t) = \alpha N(t)k_{0}a_{1} sin(\dfrac{\pi N(t)k_{0}}{2N(t)k_{0}}).
\end{equation*}
Here, $N(0)$ is the total initial population before the epidemic outbreak.

During the epidemic, as lockdown is imposed, these economically beneficial interactions also reduce. Infected individuals do not contribute to the economy; only the susceptible and recovered population is allowed to work and hence interacts with $k = k_0(1-l(t))$ other individuals that leads to value generation in the economy. Therefore, the number of economically beneficial interactions during the epidemic is given by 
\begin{equation} \label{eqn:2}
	n_{ei}(t) = \alpha N(t)k_0a_{1}sin\left(\dfrac{\pi(S(t)+R(t))k}{2N(t)k_{0}}\right).
\end{equation}
The reduced rate of interaction ($k$ instead of $k_0$) is captured in the argument of the sine function. Sine function captures the sub-linear increase in economically beneficial interaction compared to total social interaction as discussed above. The amplitude of the sine function is same as that in the pre-epidemic case.

Let the value of each economically beneficial interaction be $m_{1}$. The population also consumes at a per capita rate of $m_{2}$. The consumption is what the population spends outside the geographical area being modeled. The dynamics of the economy accounting for the value generation and consumption discussed above is given by 
\begin{equation} \label{eqn:3}
	\dot{G}(t) = m_{1}n_{ei}(t) - m_{2}N(t)
\end{equation}  

\subsection{The Optimal Control Problem}
Strict lockdown causes economic recession, whereas, leniency leads to many infections and deaths. Therefore, depending on the economic capacity of a nation and the perceived value of life, policymakers have to find out an optimal lockdown policy to balance the health and economic state of the nation. We formulate an optimal control problem that takes into account (i) the value of statistical life, (ii) the economic cost of infection (spent in treatment, long term health impairments etc.), and (iii) the economic losses due to reduced economic activities to determine the optimal lockdown policy. The objective function is given by, 
\begin{equation} \label{eqn:4}
	J = c_{1}D(T) + c_{2}(R(T)+I(T)) - G(T)
\end{equation}
Here, $D(T)$ represents total deaths by the end of the epidemic wave, and is computed in eq. \eqref{eqn:1d}. The total population who got infected during the course of the epidemic wave is given by $R(T)+I(T)$ (computed using eqns. \eqref{eqn:1b}, \eqref{eqn:1c}). The value of statistical life is given by $c_1$, and the economic cost of infection is captured by $c_{2}$. The value of the economy at the end of the epidemic wave is given by $G(T)$ (computed using eqn. \eqref{eqn:3}).

Finally, the optimal control problem can be formulated as follows:
\begin{eqnarray}
	\label{eqn:opt_control_prob}
	\min_{l(t)\in[0, l_0]} & \text{ $J$ in eqn. \eqref{eqn:4}} \\
	\text{subject to:} & \text{Epidemic state eqns. \eqref{eqn:1}} \nonumber\\
	& \text{Economic state eqn. \eqref{eqn:3}}. \nonumber
\end{eqnarray}

\section{Parameter Tuning}

In this section we tune our model with real world data for economies of various countries, and for the Covid-19 epidemic. The carrying capacity of the population is normalized to $K = 50,000$ in all cases and the economies are considered in per capita terms. We assume that the pre-epidemic population is equal to the carrying capacity, $N(0) = 50,000$  i.e. the population is in equilibrium state. The decision horizon is set to $366$ days (the epidemic wave ends in all the cases by this duration). The initial number of susceptible, infected and recovered population is set to $49500$, $500$ ($1\%$ of the total population) and $0$ respectively. The average number of interactions a person can have in absence of epidemic is set to $k_{0} = 22$ \cite{k0}.

The spreading rate of the original variant of Corona virus was estimated to be around $0.31-0.36$ for different countries \cite{beta}. We have adjusted $\beta_{0}$ such that the value of spreading rate in the uncontrolled epidemic, $\beta = \beta_{0}k_{0}$, is close to the observed value. Here, the value of $\beta $ is set to $0.33$. We assume that on an average an infected person recovers in $10$ days. Therefore, the recovery rate is $0.1$ which is close to the estimate in \cite{beta}. The infected person can die at an average death rate of $\delta = 0.004$ per day. Covid-19 fatality rate was observed to be very high in few countries like Italy, Sweden, Belgium etc. \cite{delta} in the first wave. To capture this effect, $\delta$ is adjusted such that out of all infected population (confirmed positive cases) during the course of epidemic, $3\%$ of the population dies. i.e. $D(T)/(D(T)+R(T)) \approx 0.03$. Table \ref{tab1} summarizes the epidemic parameters and the parameters that are common for all three countries. 

As discussed in Section II.B, interactions among individuals add value to the economy. Not all of the interactions generate value. Therefore, in absence of the epidemic, we assume that $60\%$ of the total interactions are economically beneficial, i.e. $a_{1} = 0.6$. We have used the data of GDP per capita in the year $2019$ for all three countries: India, US and Burundi \cite{gdp_per_capita}. Since, the pre-epidemic total population is normalized to $N(0)$ for all three countries, the GDP before epidemic, $G(0)$ is GDP per capita in the year $2019$ multiplied by $ N(0)$. Table \ref{tab2} summarizes the country specific economic parameters along with their references.

The value of statistical human life, $c_{1}$ is obtained from \cite{c1_india,c1_c2_us, c1_burundi}. These values are averaged quantities over complete lifetimes of many individuals. We assume that on an average person who has died because of the infection would have lived for $20$ more years (deaths among aged population was more during the first Covid-19 wave). Therefore, this total estimate is divided by $20$ years to get per year value of statistical life. The value of statistical human life for India, US and Burundi are $\$30,000$, $\$350,000$ and $\$467$ respectively. The cost of infection, $c_{2}$ in US is estimated to be $\$20,000$ per year per person \cite{c1_c2_us}. The numbers for Burundi and India are $\$26.7$ and $\$500$ respectively.

Per capita consumption, $m_{2}$, for India, US and Burundi (shown in Table \ref{tab2}) are obtained from \cite{m2}. These values are per capita consumption for a year so, we have divided them by $365$ to get per capita consumption per day. Now that we have value of $m_{2}$, we have adjusted the value of $m_{1}$ such that the growth in the economy is consistent with the observed data of GDP growth (normalized to population $N(0)$) for the respective countries.

\begin{table}[htbp]
	\caption{Common Parameters (Epidemic Parameters)}
	\begin{center}
		\begin{tabular}{|c|c|}
			\hline 
			\textbf{Parameter Name,} & \textbf{Parameter}\\ 
			\textbf{(Symbol)} &   \textbf{Value}\\
			\hline
			Carrying capacity of population, $K$ & 50,000\\
			\hline 
			Epidemic duration, $T$ & 366 days \\
			\hline
			Pre-epidemic avg. number of interactions, $k_{0}$ & 22 \cite{k0} \\
			\hline
			Fraction of economically beneficial interactions, $a_{1}$  & 0.6\\
			\hline
			Spreading rate, $\beta = \beta_{0}k_0$ & $0.015\times 22$ \\
			& $= 0.33$ \cite{beta}\\
			\hline
			Recovery rate, $\gamma$  & 0.1 \cite{beta}\\
			\hline
			Death rate due to infection, $\delta$ & 0.004 \\
			\hline
			Pre-epidemic population, $N(0)$ & 50,000\\
			\hline
			Initial infected population, $I(0)$ & 500\\ 
			\hline
			Initial recovered population, $R(0)$ & 0\\
			\hline
			Initial susceptible population, $S(0)$  & 49,500 \\
			\hline
		\end{tabular}
		\label{tab1}
	\end{center}
\end{table}

\begin{table}[htbp]
	\caption{Country Specific Economic Parameters}
	\begin{center}
		\begin{tabular}{|c|c|c|c|}
			\hline 
			\textbf{Parameter Name,} & \textbf{India}& \textbf{United}& \textbf{Burundi} \\ 
			\textbf{(Symbol)} & &\textbf{State}s & \\
			\hline
			Pre-epidemic GDP, & $\$2101N(0) $ & $\$65000N(0)$  & $\$261N(0)$ \\
			$G(0)$ \cite{gdp_per_capita} & & & \\
			\hline
			Employment ratio, $\alpha$ \cite{alpha} & 0.9473& 0.9633 & 0.992\\
			\hline
			Net migration rate, $\mu$ \cite{mu} & -0.000383 & 0.002893  & 0.000171  \\
			\hline
			Cost of each economically  & $\$0.2665$ & $\$13.91$ & $\$0.043015$ \\ 
			beneficial interaction, $m_{1}$ & & & \\
			\hline
			Cost of consumption, $m_{2}$ & $\$3.1$& $\$173$ & $\$0.55$ \\
			\cite{m2} & & & \\
			\hline
			Value of the statistical  & $\$30,000$& $\$350,000$  & $\$467$ \\
			life, $c_{1}$ & \cite{c1_india} &\cite{c1_c2_us} & \cite{c1_burundi} \\
			\hline
			Cost of infection, $c_{2}$  & $\$500$ & $\$20,000$ \cite{c1_c2_us} & $\$26.7$ \\
			\hline  
			
		\end{tabular}
		\label{tab2}
	\end{center}
\end{table}

\section{Numerical Solution}
The problem formulated in \eqref{eqn:opt_control_prob} is solved using the direct method of solving the optimal control problems \cite{direct_method}. We have discretized the problem using a fixed step size of $\Delta t=3$ units for the independent variable $t$. We have used standard 4th order, fixed step size, Runge-Kutta method \cite{runge_kutta} to solve the ODEs present in eqns. \eqref{eqn:1} and \eqref{eqn:3}. The chosen step size works well in practice for the range of system parameters considered in this paper. Increasing the step size makes the system unstable, on the other hand, reducing the step size increases the computation time without significantly improving the accuracy of the solution.

\section{Results}

\begin{figure*}[ht!]
	\centering
	\subfloat[Burundi \label{fig:lockdown_strength_burundi}]{
		\includegraphics[width=54mm]{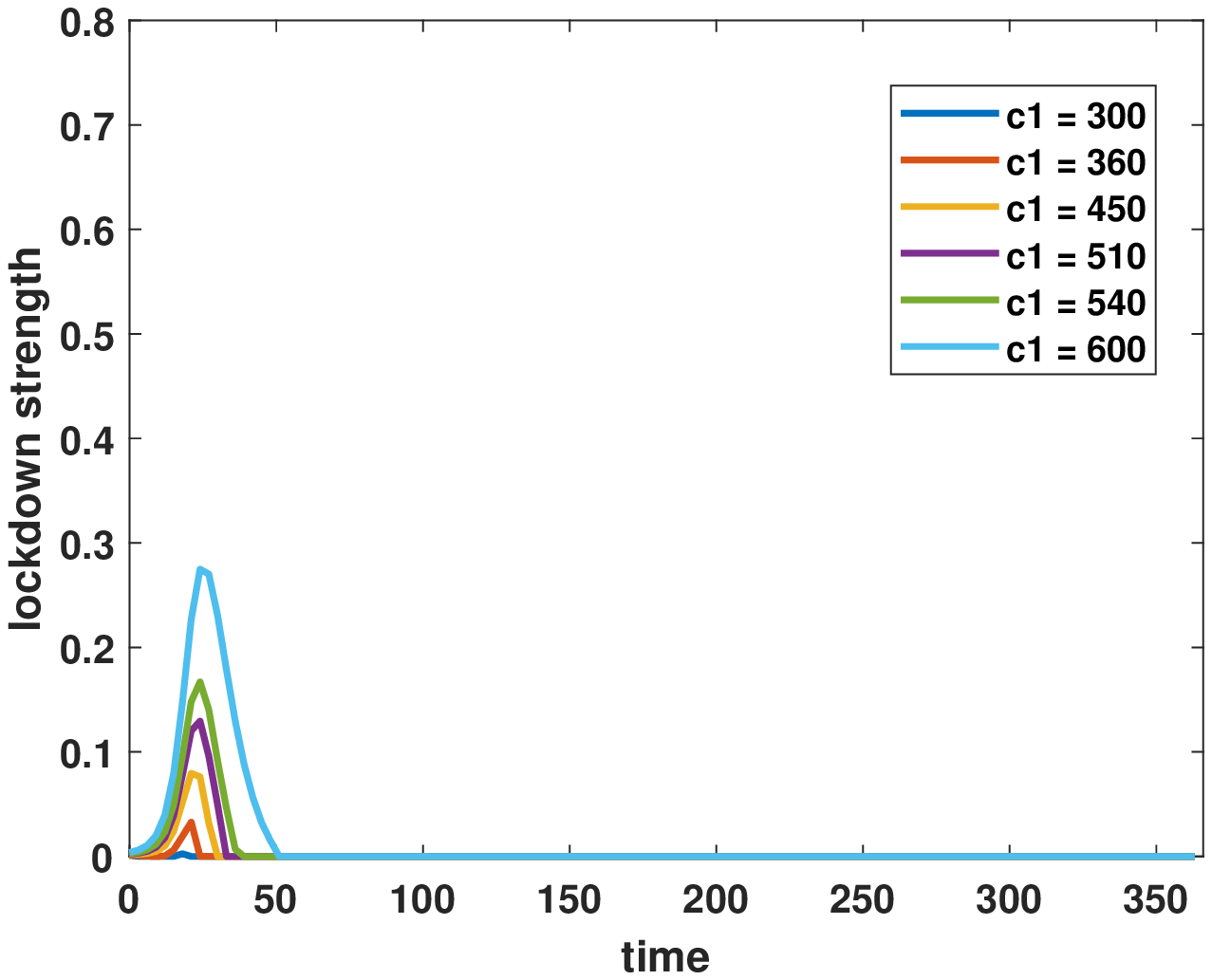} }
	\hfill
	\subfloat[India \label{fig:lockdown_strength_india}]{
		\includegraphics[width=54mm]{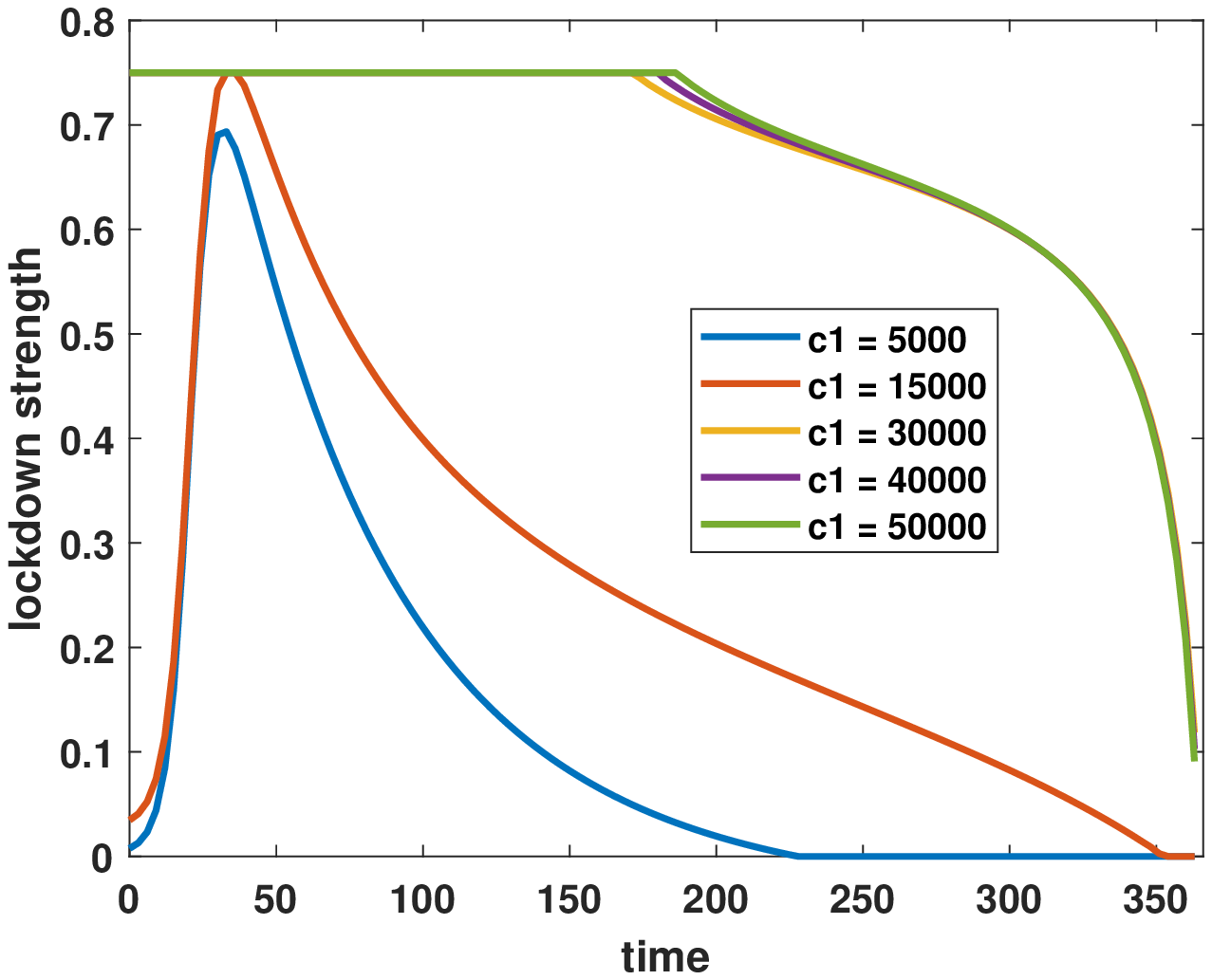} }
	\hfill
	\subfloat[United States \label{fig:lockdown_strength_us}]{
		\includegraphics[width=54mm]{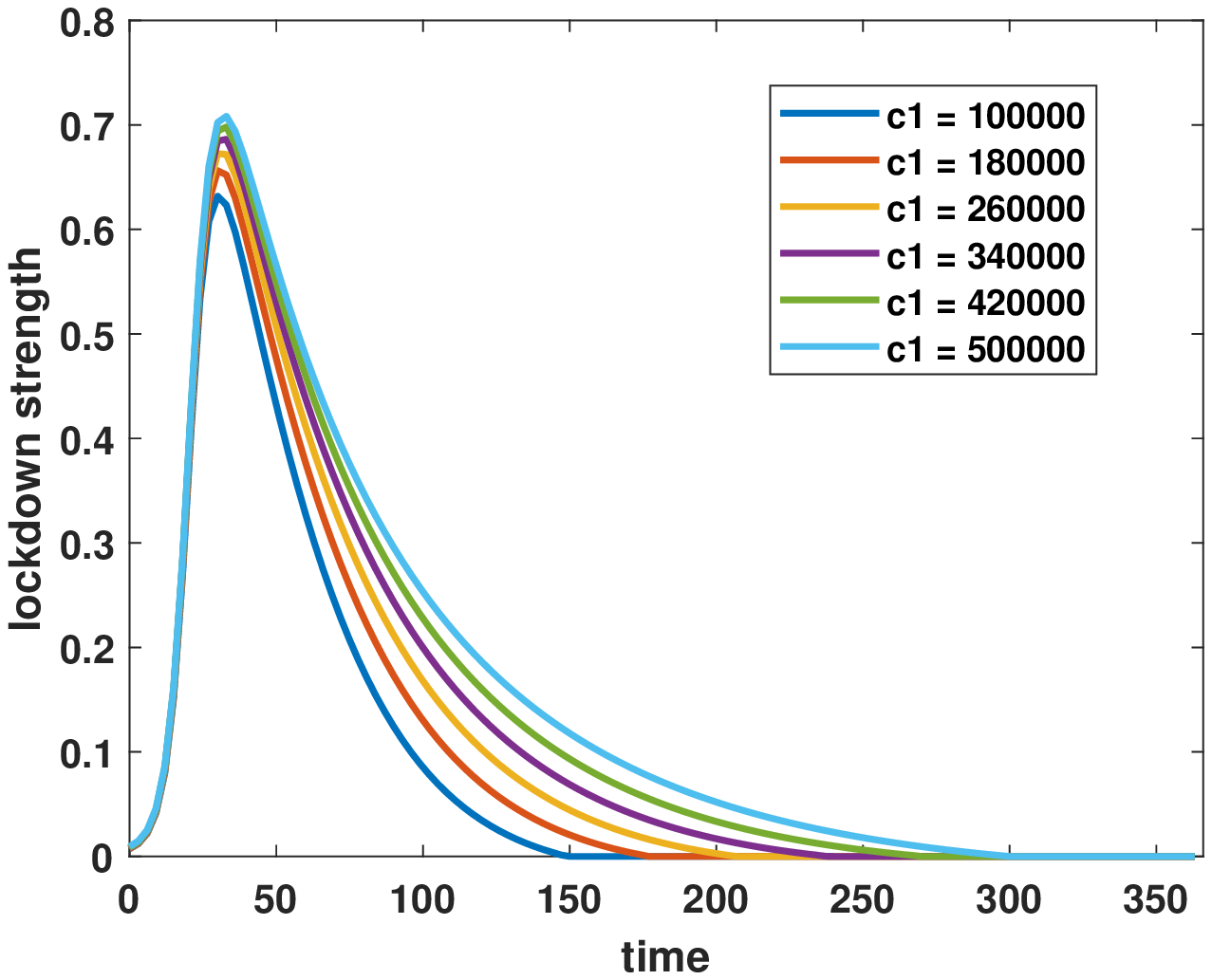} }
	\caption{\small{Lockdown strength vs. time (in days), for different values of statistical life, $c_1$.}}
	\label{fig:lockdown_strength}
\end{figure*}

\begin{figure*}[ht!]
	\centering
	\subfloat[Burundi \label{fig:components_of_cost_function_burundi}]{
		\includegraphics[width=54mm]{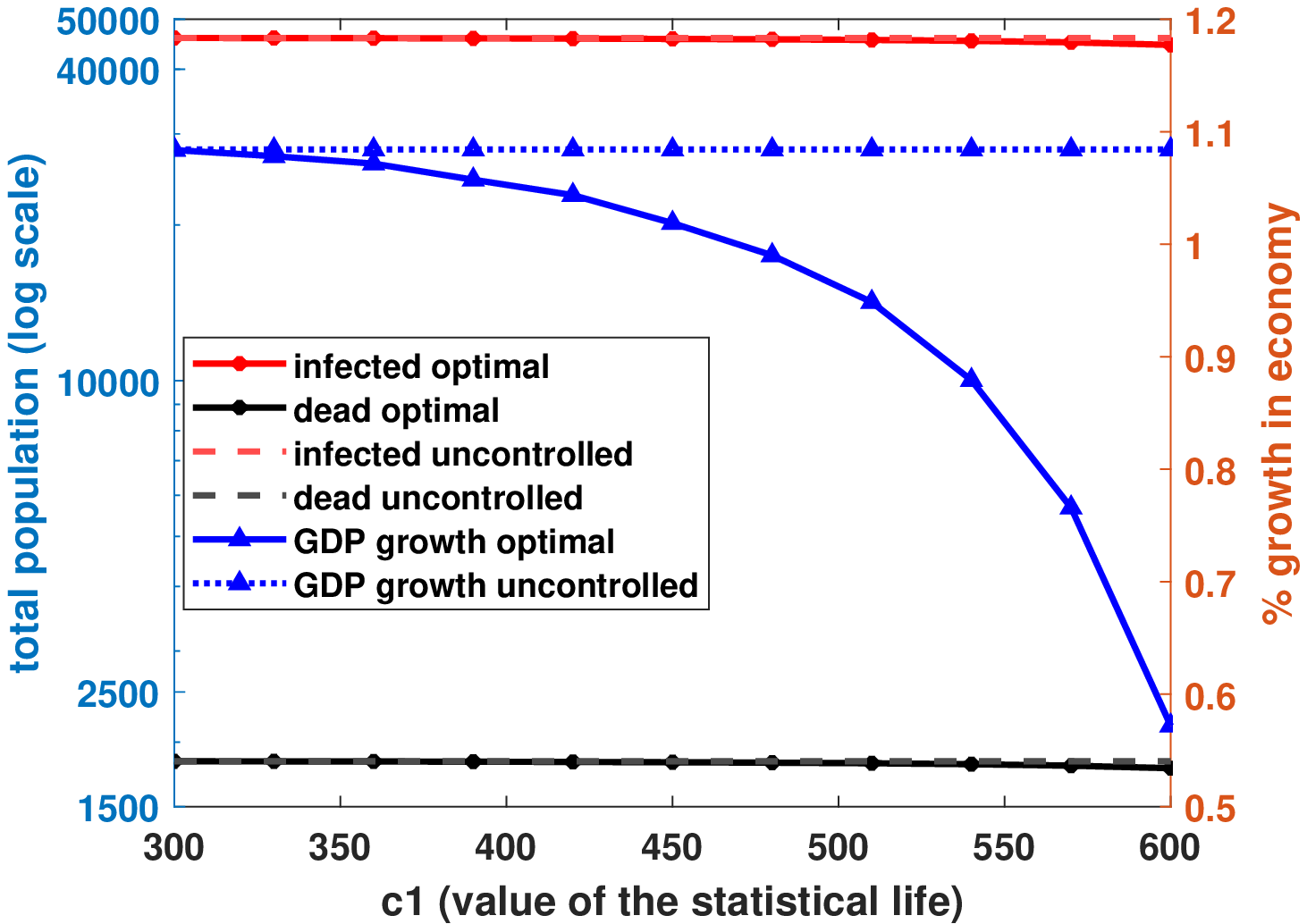} }
	\hfill
	\subfloat[India \label{fig:components_of_cost_function_india}]{
		\includegraphics[width=54mm]{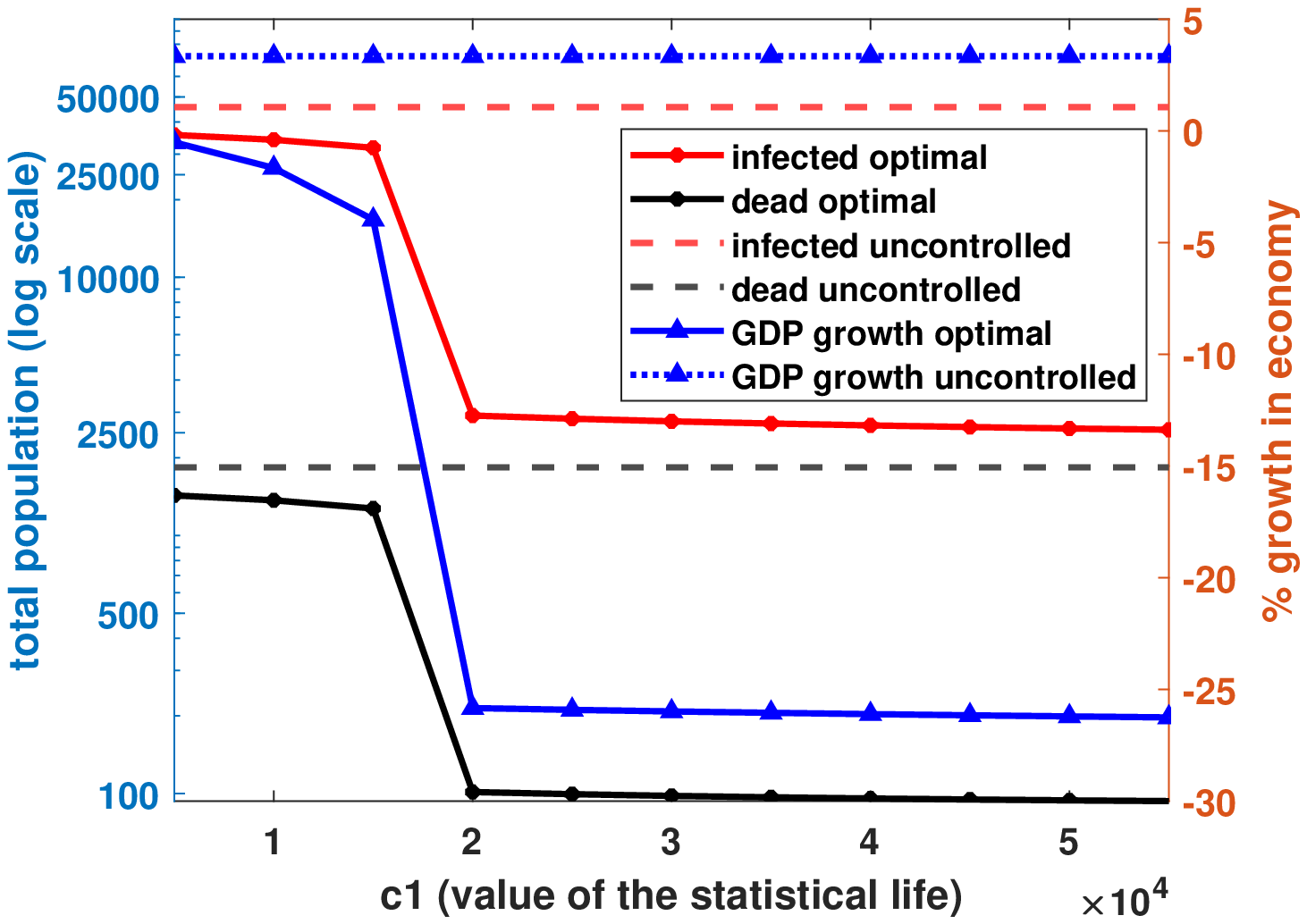} }
	\hfill
	\subfloat[United States \label{fig:components_of_cost_function_us}]{
		\includegraphics[width=54mm]{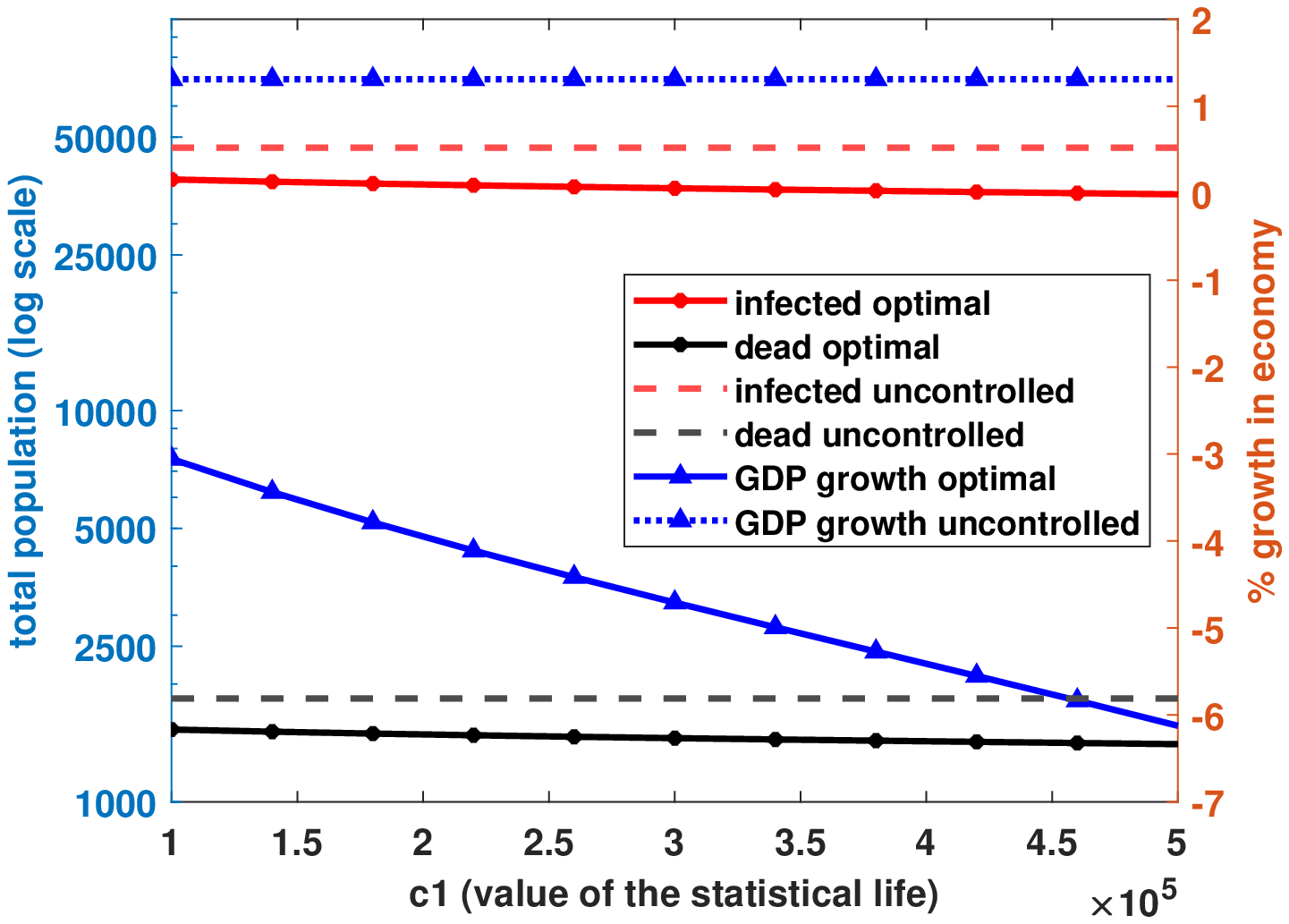} }
	\caption{\small{Components of the cost function vs. value of statistical life, $c_1$. Number of infected and dead individuals are plotted on left Y-axis and economic growth on right Y-axis for both uncontrolled and optimal cases.}}
	\label{fig:components_of_cost_function}
\end{figure*}

\begin{figure}[ht!]
	\centering
	\subfloat[India \label{fig:cost_function_india}]{
		\includegraphics[width=43mm]{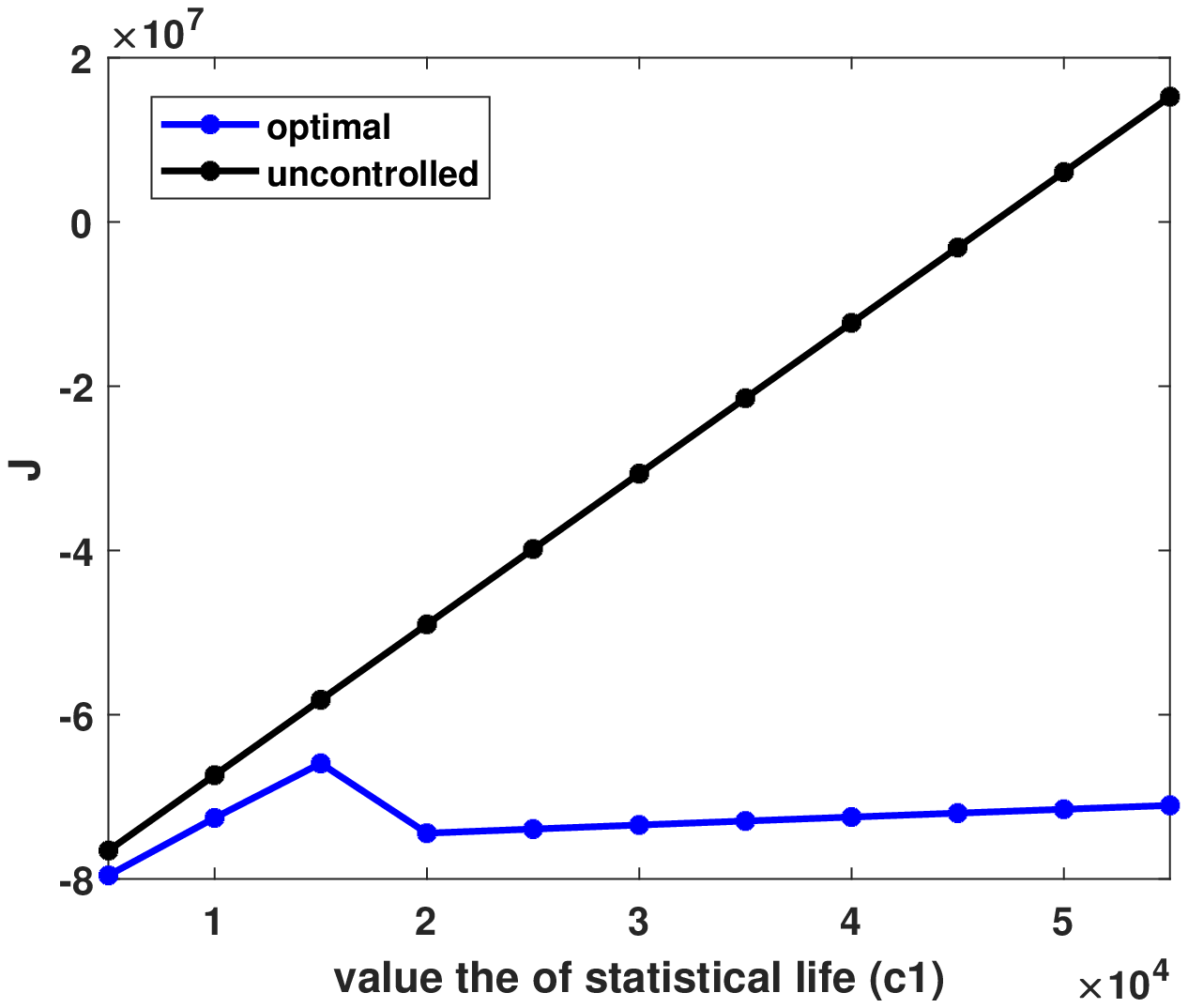} }
	\subfloat[United States \label{fig:cost_function_us}]{
		\includegraphics[width=43mm]{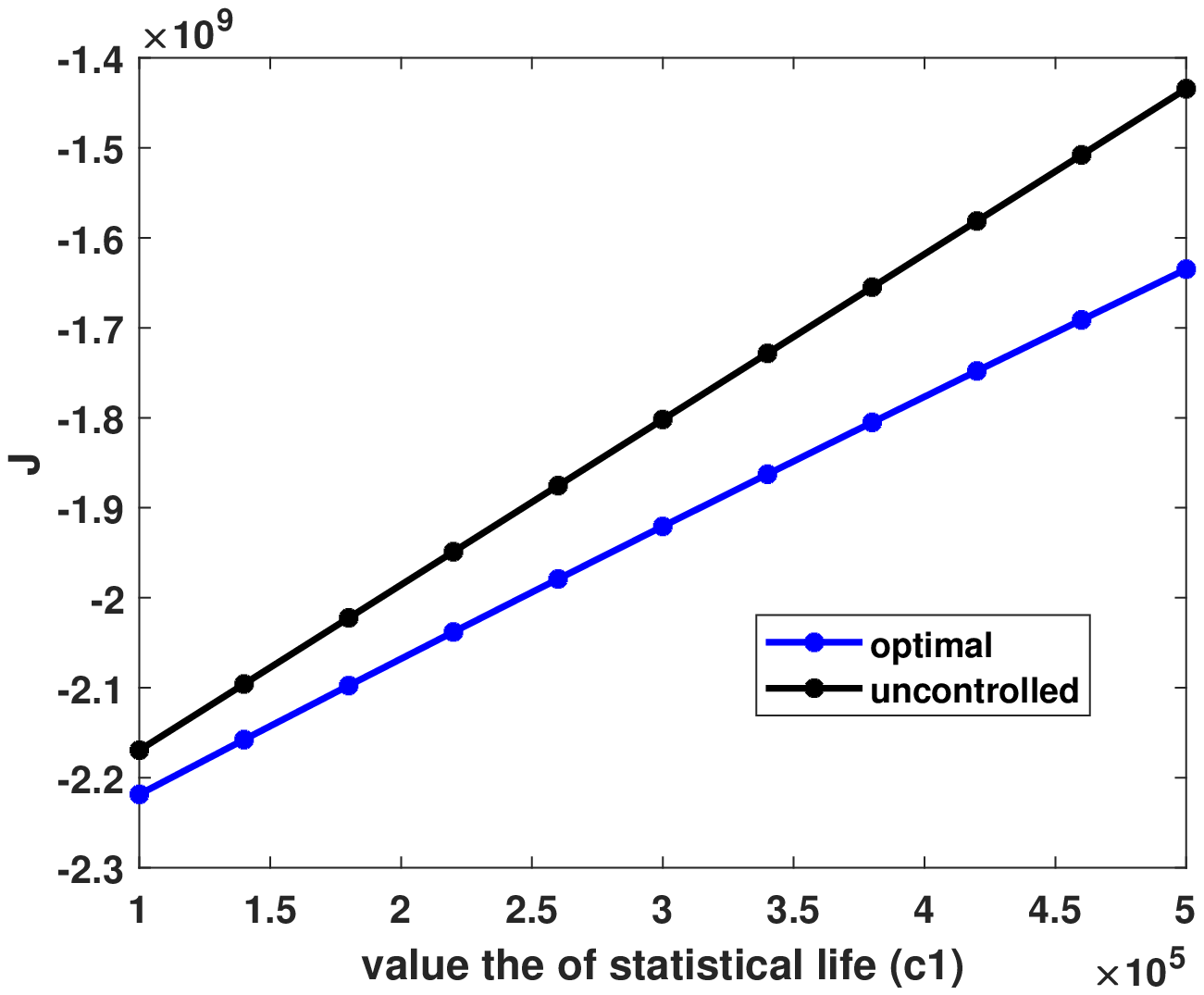} }
	\caption{\small{Cost function vs. value of statistical life, for uncontrolled and optimal cases.}}
	\label{fig:cost_function}
\end{figure}

Fig. \ref{fig:lockdown_strength} shows the strength of lockdown signal for the three countries---Burundi, India, and US---for various values of statistical life, $c_1$, calibrated according to the respective countries. Although past studies have estimated $c_1$ for various countries (Table \ref{tab2}), the ongoing Covid-19 pandemic is likely to have affected it, thus we have analyzed the optimal solution for a range of values of $c_1$ around the original estimate. 

We notice that when the economy of the country is too weak, as in the case of Burundi, the lockdown is almost non-existent (Fig. \ref{fig:lockdown_strength_burundi}). This agrees with the intuition. Locking down such an economy has extreme effects on the livelihood of individuals. If the economy of the country is very strong, as in the case of US, the lockdown imposed in the optimal case is partial (Fig. \ref{fig:lockdown_strength_us}) for a wide range of values of statistical life. The economy is too big to be ignored, hence full lockdown is avoided, however the values of life is significant enough to impose a sizable lockdown for a significant period of time. 

An economy which is somewhere in between, India, has a more interesting behavior. As $c_1$ increases, the optimal solution switches from partial lockdown to almost full lockdown. Thus, the optimal solution is more sensitive to the value of $c_1$ in this case than the previous two cases. This behavior is also in line with the observation that India had nationwide shut down during the first wave of Covid-19\footnote{For the base case parameter value $c_1=\$30,000$ for India, optimal solution is strict lockdown in Fig. \ref{fig:lockdown_strength_india}.}, but during the second wave only localized (partial) lockdown were imposed by the state governments and/or the district administrations. During both the waves, lockdown was lifted gradually.

Fig. \ref{fig:components_of_cost_function} shows the three components of the cost function $J$ defined in Eq. \eqref{eqn:4} for varying values of $c_1$. As $c_1$ increases, the economic losses increase for all three countries. For cases when extreme lockdown is imposed in India, the economic losses incurred is greater than $25\%$. It should be noted that GDP contracted by about $24\%$ in the first quarter of 2020 when the whole country was under lockdown \cite{q12020-21growth}. Stronger lockdown leads to fewer deaths and infections as can be seen in the case of India and US but at the cost of greater economic losses. The value of cost function $J$ is given in Fig. \ref{fig:cost_function} for India and US, plot for Burundi is removed due to space constraints.

\section{Conclusion}

In this paper we have formulated a problem for computing the optimal lockdown policy for an epidemic for which other mitigation strategies are unavailable. The cost function in our formulation accounts for economic losses due to lockdown as well as losses due to infections and deaths during the epidemic. We have tuned the model parameters for Covid-19 epidemic and the economic situation of low, middle and high income countries (Burundi, India, and US).

The results show that the lockdown is almost non-existent for Burundi and always partial for the US for a wide range of the value of statistical life (a measure for economic cost of each death during the epidemic). Whereas the optimal policy switches from partial lockdown to almost full lockdown as the value of statistical life increases for India. Our results are inline with the lockdown policies followed in the respective countries during the first wave of Covid-19 epidemic. The framework presented here may assist policymakers in devising optimal lockdown strategies and assessing economic losses due to them.

\end{document}